\documentclass[journal=nalefd,manuscript=letter]{achemso}

\usepackage[version=3]{mhchem} 

\usepackage{graphicx}
\usepackage{amsmath}
\usepackage{cleveref}
\usepackage{siunitx}
\usepackage{color}

\usepackage[bookmarks=true,colorlinks=true,urlcolor=blue,linkcolor=blue,citecolor=blue,breaklinks]{hyperref}

\usepackage{amssymb}
\usepackage{booktabs}

\newcommand*{\DIPC}[0]{{
Donostia International Physics Center (DIPC),
Paseo Manuel de Lardizabal 4, 20018 Donostia-San Sebasti\'an, Spain}}

\newcommand*{\FUBER}{{
Institut f{\"u}r Chemie und Biochemie, Freie Universit{\"a}t Berlin,
Takustr.~3, 14195 Berlin, Germany}}

\author{Dino Novko}
\email{dino.novko@gmail.com}
\affiliation{\DIPC}
\affiliation{\FUBER}

\title{Dopant-induced plasmon decay in graphene}


\begin{document}


\begin{abstract}


Chemically doped graphene could support plasmon excitations up to telecommunication or even visible frequencies. Apart from that, the presence of dopant may influence electron scattering mechanisms in graphene and thus impact the plasmon decay rate. Here I study from first principles these effects in single-layer and bilayer graphene doped with various alkali and alkaline earth metals. I find new dopant-activated damping channels: loss due to out-of-plane graphene and in-plane dopant vibrations, and electron transitions between graphene and dopant states. The latter excitations interact with the graphene plasmon and together they form a new hybrid mode. The study points out a strong dependence of these features on the type of dopants and the number of layers, which could be used as a tuning mechanism in future graphene-based plasmonic devices.


\end{abstract}


Recently, the quantized collective motion of surface electrons, called surface plasmon, has gained renewed attention as the potential mechanism for the confinement of electromagnetic energy, which could reduce the size of optical devices to the desired nanoscale~\cite{bib:west10}. The two-dimensional (2D) plasmon of graphene is a most promising framework to investigate these confinement effects~\cite{bib:abajo14}, as a result of its relatively long lifetime~\cite{bib:jablan09,bib:yan13,bib:woessner15} and its tunability through an electrostatic gating~\cite{bib:li08,bib:fei12} or chemical doping~\cite{bib:fedorov14,bib:vito17,bib:shirodkar17}. Angle-resolved photoemission (ARPES) studies show that chemical doping by deposition of alkali and alkaline earth metal (X) atoms on graphene introduces much higher concentrations of conducting electrons than the standard electrostatic gating techniques~\cite{bib:fedorov14,bib:mcchesney10,bib:petrovic13}. In fact, two recent theoretical studies point out that lithium-doped single- and few-layer graphene can support plasmons ranging from near-infrared to possibly visible energies due to a high level of doping~\cite{bib:vito17,bib:shirodkar17}. This opens new possibilities to extend the application of graphene plasmonics to telecommunication technologies~\cite{bib:west10,bib:abajo14}, photodetectors~\cite{bib:liu14}, or photovoltaic systems~\cite{bib:park14}.

The underlying physics of dopant-induced plasmon decay in graphene, i.e., how dopants affect the electron scattering processes, is not understood yet. The largest contribution consists of interband electron-hole pair excitations between occupied and unoccupied $\pi$ bands~\cite{bib:jablan09} (i.e., Landau damping), which are suppressed due to Pauli blocking below the value of two times the Fermi energy, $2\varepsilon_F$. Since the value of $\varepsilon_F$ in X-doped graphene shifts up to $\sim1.5\,\mathrm{eV}$~\cite{bib:mcchesney10,bib:petrovic13,bib:fedorov14}, this damping channel is diminished within a large energy window. Nevertheless, the 2D plasmon in doped graphene can still show substantial decay rates below the interband gap because of higher-order processes: electron-phonon~\cite{bib:jablan09,bib:carbotte10,bib:stauber08,bib:yan13}, electron-impurity~\cite{bib:principi13a}, and electron-electron~\cite{bib:principi13b,bib:grushin09} scatterings. For the case of dopant-free graphene it is widely accepted that the first decay channel is a major contributor to the plasmon decay rate, but only when the plasmon energy exceeds the energy of intrinsic optical phonon of graphene ($\omega_{\mathrm{op}}\approx0.2\,\mathrm{eV}$)~\cite{bib:jablan09,bib:yan13}. On the other hand, when the plasmon energies are below this value, the main sources of damping are the latter two scattering channels~\cite{bib:principi14,bib:woessner15}. ARPES studies have shown a large increase in the electron-phonon interaction of graphene when X atoms are present~\cite{bib:fedorov14,bib:haberer13}, and that the magnitude of the increase depends strongly on the dopant species~\cite{bib:fedorov14}. As one of the consequences, the highly doped graphene is believed to be a promising material for achieving phonon-mediated superconductivity in 2D systems~\cite{bib:profeta12,bib:yang14}. However, the needed decisive quantitative analyses of the 2D plasmon and its decay rate are still lacking in this case.

In this Letter I report the phonon-induced plasmon decay in single-layer and bilayer graphene doped with various X atoms, where X = Li, Ca, Sr, K, Cs and Ba. By using \textit{ab initio} based methods I show that the presence of X atoms (i) alters significantly the electron-phonon coupling strength of intrinsic optical phonon in graphene and as well (ii) introduces new phonon modes that allow plasmon decay even below 0.2\,eV. In accordance with the experiments, the calculations show that Li, Ca, and Sr produce the largest electron-phonon coupling strengths, while K, Cs, and Ba the smallest. Furthermore, due to $\sigma$ states contributed by the X atoms a new decay channel is open, namely, the low-energy interband transitions between $\pi$ and $\sigma$ states. Besides, the interaction between these interband excitations and the 2D plasmon produces a new hybrid mode characterized by the anticrossing splitting in the vicinity of their energies. Interestingly, both plasmon decay rate and dispersion relation of this hybrid mode depend strongly on the type of X atom and the number of layers, leading to new possibilities for plasmon tuning in graphene.

The central quantity for understanding scattering processes of charge carriers in materials is the optical conductivity
$
\sigma_{\alpha\alpha}(\omega)=\frac{i}{\omega}\pi_{\alpha\alpha}(\omega),
$
~\cite{bib:allen71}
where $\pi_{\alpha\alpha}$ is the current-current correlation function~\cite{bib:novko16}, and $\alpha$ is the polarization direction. 
For detailed analysis of different scattering channels it is useful to distinguish the intraband (i.e., Drude) from the interband transitions, i.e., $\sigma_{\alpha\alpha}=\sigma^{\mathrm{d}}_{\alpha\alpha}+\sigma^{\mathrm{ib}}_{\alpha\alpha}$. 

The electron-phonon scattering mechanism is incorporated into $\sigma^{\mathrm{d}}_{\alpha\alpha}$ by using the Holstein theory for normal metals~\cite{bib:holstein64}. Within this theory the conductivity is calculated by means of a diagrammatic analysis and solving the Bethe-Salpeter equation for $\pi^{\mathrm{d}}_{\alpha\alpha}$~\cite{bib:kupcic15}. According to Allen~\cite{bib:allen71}, the solution can be written as the following
\begin{eqnarray}
\label{eq2}
\pi^{\mathrm{d}}_{\alpha\alpha}(\omega)&=&\frac{2}{\Omega}\sum_{\mathbf{k},n}\left(-\frac{\partial f_{n\mathbf{k}}}{\partial\varepsilon_{n\mathbf{k}}}\right)\left|j^{\alpha}_{nn\mathbf{K}}\right|^2\nonumber\\
&&\times\frac{\omega}{\omega\left[1+\lambda_{\mathrm{ph}}(\omega)\right]+i/\tau_{\mathrm{ph}}(\omega)},
\end{eqnarray}
where $\varepsilon_{n\mathbf{k}}$ are single-electron energies, $f_{n\mathbf{k}}$ is the temperature-dependent Fermi-Dirac distribution function, $j^{\alpha}_{nm\mathbf{K}}$ are the current vertices, and $\Omega$ is the normalization volume. The effects of the electron-phonon interaction are contained in the dynamical renormalization  and scattering time parameters, i.e., $\lambda_{\mathrm{ph}}(\omega)$ and $\tau_{\mathrm{ph}}(\omega)$, respectively~\cite{bib:allen71,bib:kupcic15,bib:kupcic17}. These two quantities are related by the Kramers-Kronig relations. In the low-temperature regime ($k_BT\ll\omega$) $\tau_{\mathrm{ph}}(\omega)$ can be written as
\begin{eqnarray}
\label{eq3}
1/\tau_{\mathrm{ph}}(\omega)=\frac{2\pi}{\omega}\int_0^{\omega}d\omega'(\omega-\omega')\alpha^2F(\omega').
\end{eqnarray}
The Eliashberg function $\alpha^2F(\omega)$ contains the information on phonon density of states and electron-phonon coupling strengths. Equation \eqref{eq2} describes the indirect intraband phonon-assisted processes in absorption spectra beyond the usual second-order perturbation theory~\cite{bib:allen71,bib:brown16,bib:giustino17,bib:novko16a}, which means that it is valid, in contrast to the latter theory, also in the low-energy regime. In order to account for other scattering mechanisms, e.g., electron-impurity and electron-electron scatterings, a constant phenomenological damping rate is added to $1/\tau_{\mathrm{ph}}(\omega)$, i.e., $1/\tau_{\mathrm{ph}}(\omega)\rightarrow1/\tau_{\mathrm{ph}}(\omega)+1/\tau$. For the sake of comparison the same constant value of $1/\tau=5\,\mathrm{meV}$ is added in each of the studied cases. This does not affect the conclusions regarding the effects of electron-phonon damping channel.

The interband part can be written as~\cite{bib:novko16}
\begin{eqnarray}
\pi^{\mathrm{ib}}_{\alpha\alpha}(\omega)=\frac{2}{\Omega}\sum_{\mathbf{k},n\neq m}\frac{\omega\left|j^{\alpha}_{nm\mathbf{k}}\right|^2}{\varepsilon_{m\mathbf{k}}-\varepsilon_{n\mathbf{k}}}\,\frac{f_{n\mathbf{k}}-f_{m\mathbf{k}}}{\omega+\varepsilon_{n\mathbf{k}}-\varepsilon_{m\mathbf{k}}+i/\tau_{\mathrm{ib}}}.
\label{eq4}
\end{eqnarray}
Here a constant interband relaxation rate $1/\tau_{\mathrm{ib}}$ is introduced (i.e., $30\,\mathrm{meV}$ in each of the studied cases) and the indirect interband phonon-assisted processes~\cite{bib:chakraborty78} are not included. As will be seen shortly, the interband transitions have a strong impact on decay rate and dispersion relation of the 2D plasmon in X-doped graphene.

The Kohn-Sham energies and wavefunctions needed for constructing Eqs.\,\eqref{eq2}--\eqref{eq4} are calculated using the plane-wave based {\sc quantum espresso} (QE) package~\cite{bib:qe} (see SI for more information on the ground state calculations). The electron momentum summations in Eqs.\,\eqref{eq2} and \eqref{eq4} are done on a $(400\times400\times1)$ Monkhorst-Pack grid including up to 20 unoccupied electronic bands. The current vertices are calculated as in Ref.\,\citenum{bib:novko16}. Phonon-related properties (phonon energies and electron-phonon matrix elements) are obtained using density functional perturbation theory~\cite{bib:baroni01} as implemented in QE. $\alpha^2F(\omega)$ is calculated on $(200\times200\times1)$ and $(20\times20\times1)$ electron and phonon momentum grids, respectively.

The excitation spectra including plasmon modes are obtained by screening $\pi_{\alpha\alpha}$ with the help of the random phase approximation based in the density functional theory~\cite{bib:despoja13,bib:novko15,bib:novko16}. The modes are analyzed by taking the imaginary part of the screened $\pi_{\alpha\alpha}$, i.e. $A(\mathrm{q},\omega)\propto\mathrm{Im}\,\widetilde{\pi}_{\alpha\alpha}(\mathrm{q},\omega)/\omega$, where the $\mathrm{q}$-dependence comes from the 2D Coulomb potential (for more details see Ref.\,\citenum{bib:novko16}).
Since the supercell approach is used, the crucial step in obtaining the correct 2D excitation spectrum is to prevent the Coulomb interaction between the repeated slabs in the normal direction. This is achieved by integrating the Dyson equation for $\pi_{\alpha\alpha}$ only between the normal direction boundaries of the corresponding 2D slab~\cite{bib:despoja13,bib:novko15,bib:pisarra16,bib:novko16,bib:gomez16,bib:vito17}.

The presence of the X dopants introduces high concentrations of conducting electrons and thus shifts $\varepsilon_F$ away from the Dirac point, $\varepsilon_D$. This enables strong electron-phonon coupling in the intraband channel, which is otherwise suppressed in the pristine graphene. The electronic structure calculations of single-layer graphene (SLG) give the largest value of $\varepsilon_F$ for $\mathrm{X=Li}$ ($\varepsilon_F\approx1.6$\,eV), and the smallest one for $\mathrm{X=Cs}$ ($\varepsilon_F\approx1.1$\,eV), which is in a good agreement with the experiment~\cite{bib:fedorov14}. This suggests that the strongest (weakest) phonon-induced damping should be obtained for the former (latter) case.

\begin{figure}[b]
\includegraphics[width=0.5\textwidth]{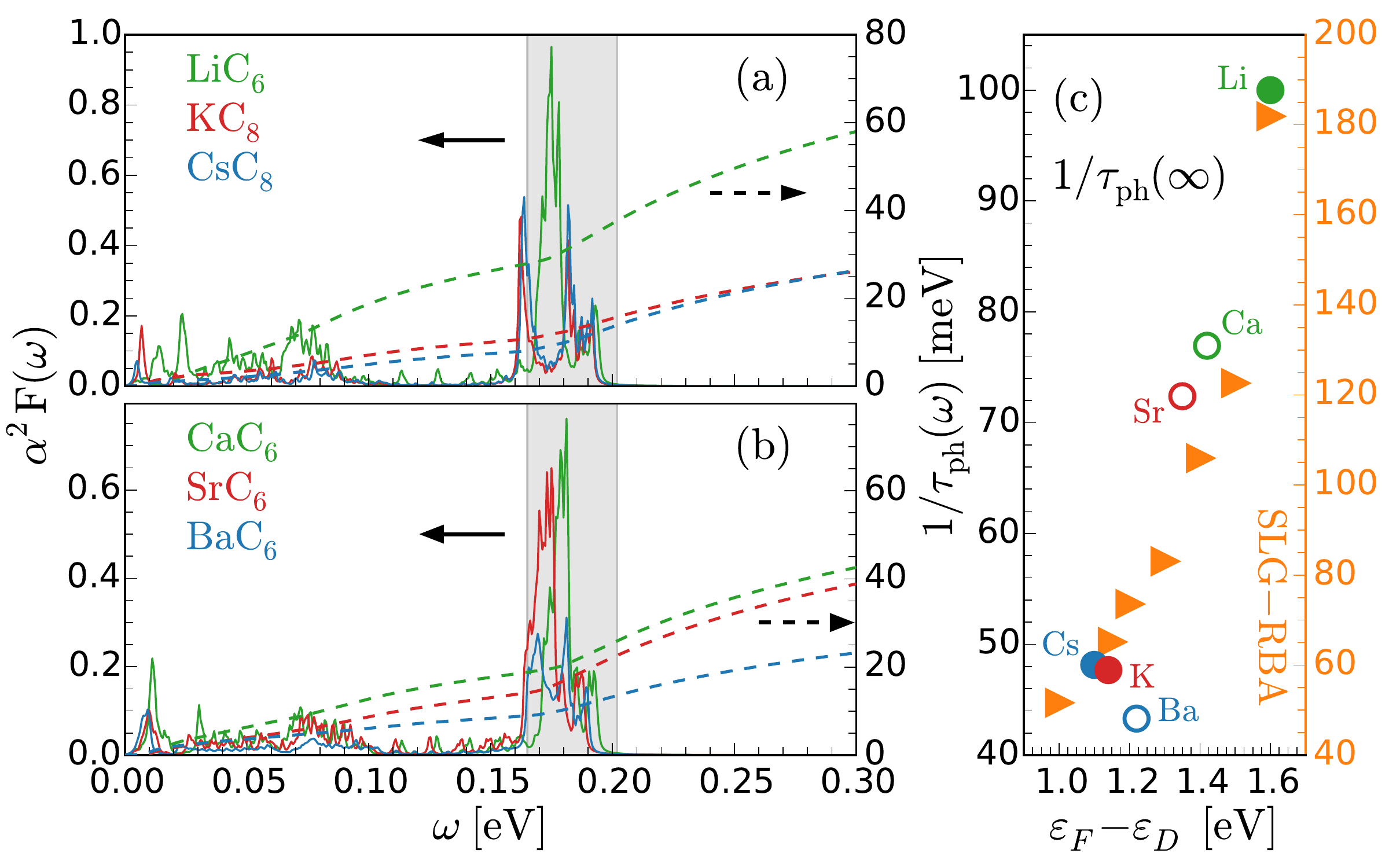}
\caption{\label{fig:fig1}(a),\,(b) The Eliashberg function $\alpha^2F(\omega)$ (left y-axis) and the damping rate due to electron-phonon coupling $1/\tau_{\mathrm{ph}}(\omega)$ (right y-axis) for graphene doped with different X atoms. Grey shaded areas represent the energy window of the intrinsic in-plane optical phonon band of dopant-free graphene. (c) High-energy limit ($\omega\rightarrow\infty$) of $1/\tau_{\mathrm{ph}}(\omega)$ for X-doped graphene as a function of energy difference between Fermi level and Dirac point (left y-axis). Orange triangles (right y-axis) are the corresponding results for SLG with a rigidly shifted Fermi energy (RBA, see text).}
\end{figure}

In Figs.\,\ref{fig:fig1}(a) and \ref{fig:fig1}(b) I show $\alpha^2F(\omega)$ and $1/\tau_{\mathrm{ph}}(\omega)$ for X-doped SLG. The corresponding stoichiometries are assumed to be as in the bulk (graphite) systems~\cite{bib:dresselhaus02}, i.e., XC$_6$ for $\mathrm{X=Li}$, Ca, Sr, and Ba [$p(\sqrt{3}\times\sqrt{3})-R\ang{30}$ dopant pattern with respect to graphene], while XC$_8$ for $\mathrm{X=K}$ and Cs [$p(2\times2)$ dopant pattern with respect to graphene] (see SI for further details). Hitherto the literature has focused in the intrinsic in-plane optical or acoustic phonons of graphene~\cite{bib:jablan09,bib:shirodkar17}. However, here we see that this simplified picture is drastically modified when the X dopants are present. Namely, (i) the electron-phonon coupling strengths of the intrinsic in-plane optical phonons are different for different dopants and (ii) new low-energy phonon modes contribute to the phonon-induced damping.

The feature (i) can be observed in the range $\omega=0.165-0.195$\,eV in $\alpha^2F(\omega)$ for different X dopants. The smallest (largest) intensities are obtained for Ba and Cs (Ca and Li). It can be also observed that the presence of X atoms shifts the upper boundary of this phonon band from 0.202\,eV found in dopant-free graphene to 0.195\,eV (see also SI). This is a consequence of the Kohn anomaly coming from electron doping, which softens the intrinsic in-plane optical modes around the $\overline{\Gamma}$ point of the Brillouin zone~\cite{bib:lazzeri06}. The feature (ii) is manifested as the appearance of the spectral peaks in $\alpha^2F(\omega)$ below the energies of the intrinsic in-plane optical phonons of graphene. These new contributions are due to the coupling with the vibrations of the X atoms and the out-of-plane graphene modes (see below). It can be seen that Li and Ca dopants give rise to the largest intensities of these low-energy peaks. As a result of (i) and (ii) a different energy dependence of phonon-induced damping $1/\tau_{\mathrm{ph}}(\omega)$ is obtained for different dopants.

The joint effects of (i) and (ii) can be also seen in the phonon-induced damping at very large energies, i.e., the limit $1/\tau_{\mathrm{ph}}(\infty)$. From Fig.\,\ref{fig:fig1}(c) it is seen that Li, Ca, and Sr dopants give the largest, while Ba, K, and Cs the smallest values of $1/\tau_{\mathrm{ph}}(\infty)$. Note that $1/\tau_{\mathrm{ph}}(\infty)\approx\pi\left\langle \omega\lambda_{\mathrm{ph}}(0) \right\rangle$~\cite{bib:allen71}, where $\lambda_{\mathrm{ph}}(0)$ is the standard electron-phonon coupling constant and $\left\langle\dots\right\rangle$ stands for the summation over all $\mathbf{q}$ and phonon bands. This relation allows a qualitative comparison with the experimental results of $\lambda_{\mathrm{ph}}(0)$. ARPES measurements yield the largest (smallest) values of $\lambda_{\mathrm{ph}}(0)$ for Ca and Li (Cs and K)~\cite{bib:fedorov14}, which is in a good qualitative agreement with the results for $1/\tau_{\mathrm{ph}}(\infty)$ presented here~\bibnote{Note that the experiments obtain larger electron-phonon coupling for Ca than Li. The reason could be that Ca atoms are actually intercalated between the graphene sheet and Au substrate, which could significantly alter the values of $\lambda_{\mathrm{ph}}(0)$~\cite{bib:fedorov14}.}. This ordering is the consequence of different factors, such as the density of states at $\varepsilon_F$, $N(\varepsilon_F)$, the mass of the X atom, $M$, the characteristic phonon frequency, $\omega_{\mathrm{ph}}$, and the deformation potential, $D$~\cite{bib:profeta12,bib:fedorov14}. In fact, using a crude approximation, one can write $\left\langle \omega\lambda_{\mathrm{ph}}(0) \right\rangle\approx N(\varepsilon_F)D^2/M\omega_{\mathrm{ph}}$. For example, Li has the smallest $M$ and is adsorbed closest to graphene layer, which introduces the largest deformation potential $D$, and thus largest $1/\tau_{\mathrm{ph}}(\infty)$~\cite{bib:profeta12} (see SI for more information). When comparing these results with those for SLG with a rigidly shifted Fermi energy (i.e., rigid-band approximation, RBA), it is seen that the actual dopants induce much smaller values of the phonon-induced plasmon damping. In order to clarify this, I note that electron doping is introduced in the RBA only by shifting the value of $\varepsilon_F$ in the Fermi-Dirac distribution functions entering Eqs.\,\eqref{eq2}--\eqref{eq4}. In other words, the RBA does not account for the changes in the phonon density of states and, importantly, in the deformation potential $D$, otherwise present in more realistic doping scenarios~\cite{bib:margine14}, and therefore overestimates the electron-phonon coupling strengths. On the contrary, X dopants are not only providing additional scattering channels, but are also considerably modifying $D$ and phonon density of states on the adiabatic level due to the presence of the excess conducting electrons. The overall outcome of these significant changes introduced by X dopants is thus underlined in the different values of $1/\tau_{\mathrm{ph}}(\infty)$ obtained in these two cases. Nevertheless, in both cases $1/\tau_{\mathrm{ph}}(\infty)$ has a qualitatively similar dependence on the position of $\varepsilon_F$.

To emphasize the role of the different phonon modes in the plasmon damping it is useful to decompose $1/\tau_{\mathrm{ph}}(\infty)$ in C and X atoms in-plane ($xy$) and out-of-plane ($z$) contributions. Table\,\ref{tab:table1} shows that the largest contribution to $1/\tau_{\mathrm{ph}}(\infty)$ is due to the C$_{xy}$ modes. However, considerable role in the plasmon damping is played by the X$_{xy}$ and C$_z$ modes (see SI for more information). The latter mode couples weakly with electrons in dopant-free SLG, but in X-doped SLG, due to the presence of $\sigma$ states, the coupling is enhanced by the interband $\pi\rightarrow\sigma$ scatterings~\cite{bib:profeta12}. Here it is again seen that the RBA largely overestimates the C$_{xy}$ contribution.

\begin{figure}[b]
\includegraphics[width=0.5\textwidth]{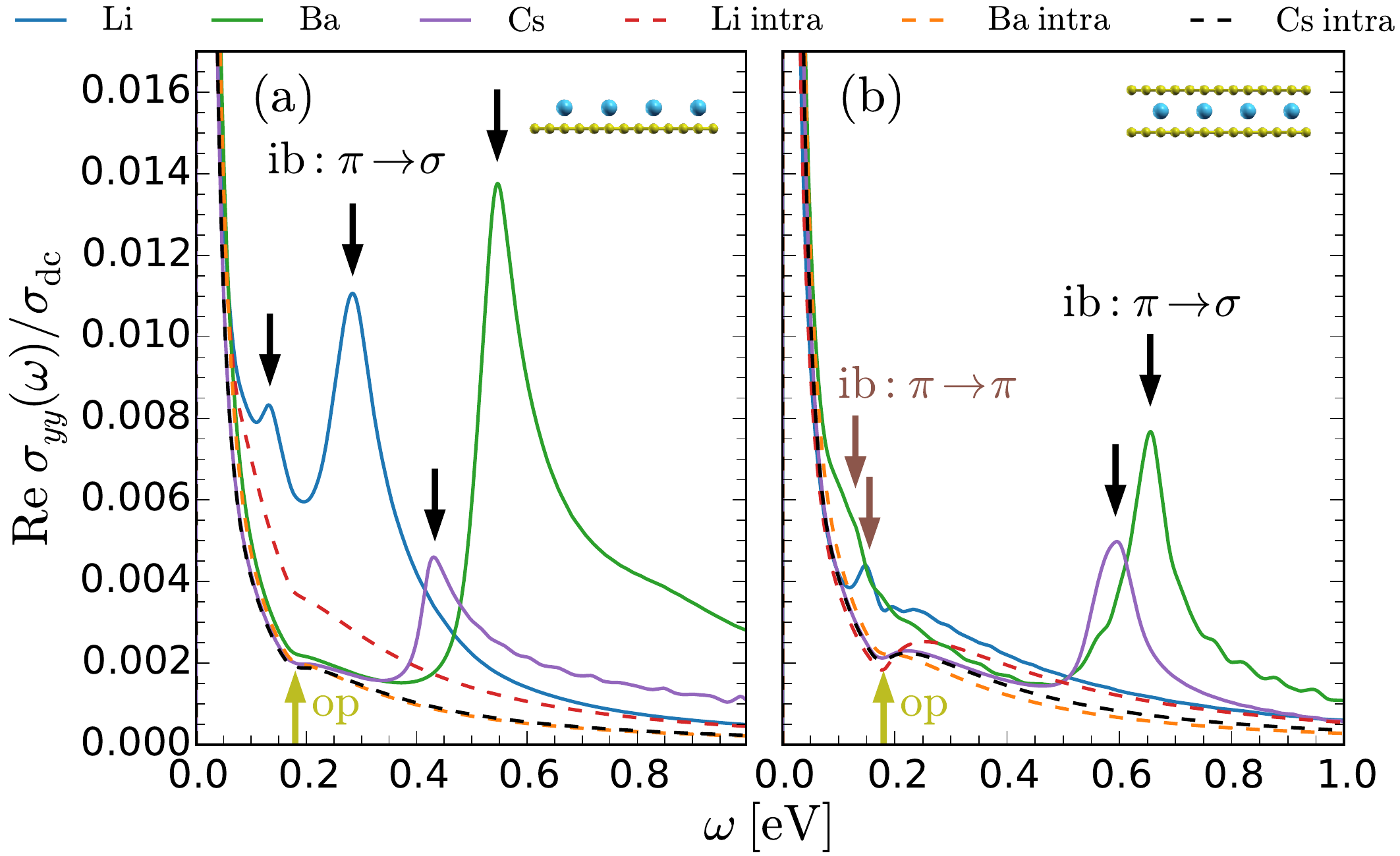}
\caption{\label{fig:fig2}Optical absorption $\mathrm{Re}\,\sigma_{yy}(\omega)$ normalized to $\sigma_{yy}(0)\equiv\sigma_{\mathrm{dc}}$ for X-doped (a) single-layer graphene and (b) bilayer graphene. Green arrows indicate the approximate energy of the C$_{xy}$ optical phonon (op) band. Black and brown arrows indicate the energy peak positions of the interband (ib) $\pi\rightarrow\sigma$ (i.e., from graphene to X atom) and $\pi\rightarrow\pi$ (i.e., from the one graphene layer to the other) transitions, respectively.} 
\end{figure}

The overall plasmon decay rate, consisting of phonon-induced and Landau dampings, can be extracted from the optical absorption spectra, since $1/\tau_{\mathrm{tot}}\propto \mathrm{Re}\,\sigma_{\alpha\alpha}$ holds~\cite{bib:kupcic15,bib:brown16}. Therefore, a large absorption intensity at a given energy implies large plasmon damping at the same energy. Figure~\ref{fig:fig2}(a) shows $\mathrm{Re}\,\sigma_{yy}(\omega)$ for SLG with $\mathrm{X=Li}$, Ba, and Cs normalized to its direct current (static) value $\sigma_{\mathrm{dc}}$. The results for the intraband channel (dashed lines) show an increase in $\mathrm{Re}\,\sigma^{\mathrm{d}}_{yy}(\omega)$ around the energies of the C$_{xy}$ optical phonon band. For $\mathrm{X=Li}$ the increase in $\mathrm{Re}\,\sigma^{\mathrm{d}}_{yy}(\omega)$ appears even below these energies due to the significant role of the C$_{z}$ and X$_{xy}$ modes in this case (dashed red line). The comparison of the total optical absorption $\mathrm{Re}\,\sigma_{yy}(\omega)$ with $\mathrm{Re}\,\sigma^{\mathrm{d}}_{yy}(\omega)$ reveals the emergence of a new interband channel that introduces large optical absorption and hence large $1/\tau_{\mathrm{tot}}$. The usual interband $\pi\rightarrow\pi$ transitions are suppressed here up to energies of $\sim3$\,eV, and the new low-energy channel originates from the $\pi\rightarrow\sigma$ (i.e., from graphene to X atom) transitions (black arrows). For the sake of completeness, the results for the X-doped bilayer graphene, where the X atoms are intercalated between two graphene layers, are also shown [see Fig.\,\ref{fig:fig2}(b)]. In that case, the peaks of the $\pi\rightarrow\sigma$ transitions are pushed towards higher energies (i.e., the $\sigma$ band is less occupied)~\bibnote{Another possibility leading to the shift of the $\pi\rightarrow\sigma$ damping channel to higher energies would be to change the coverage of X atoms~\cite{bib:vito17}}. In fact, for $\mathrm{X=Li}$ this damping channel is shifted to $\sim2$\,eV (not shown). For bilayer systems an additional plasmon damping channel appears at low energies: the interband $\pi\rightarrow\pi$ (i.e., between graphene layers) transition (brown arrows). The intensity of the latter is, however, much smaller than the intensity of the $\pi\rightarrow\sigma$ channel, and thus it contributes less to $1/\tau_{\mathrm{tot}}$. These new interband excitations, especially the prominent $\pi\rightarrow\sigma$ transitions, should be experimentally observable through infrared optical spectroscopy, in a similar manner as the low-energy interband excitations were measured in dopant-free bilayer graphene~\cite{bib:wang08,bib:li09}.

\begin{figure*}[!htbp]
\includegraphics[width=0.95\textwidth]{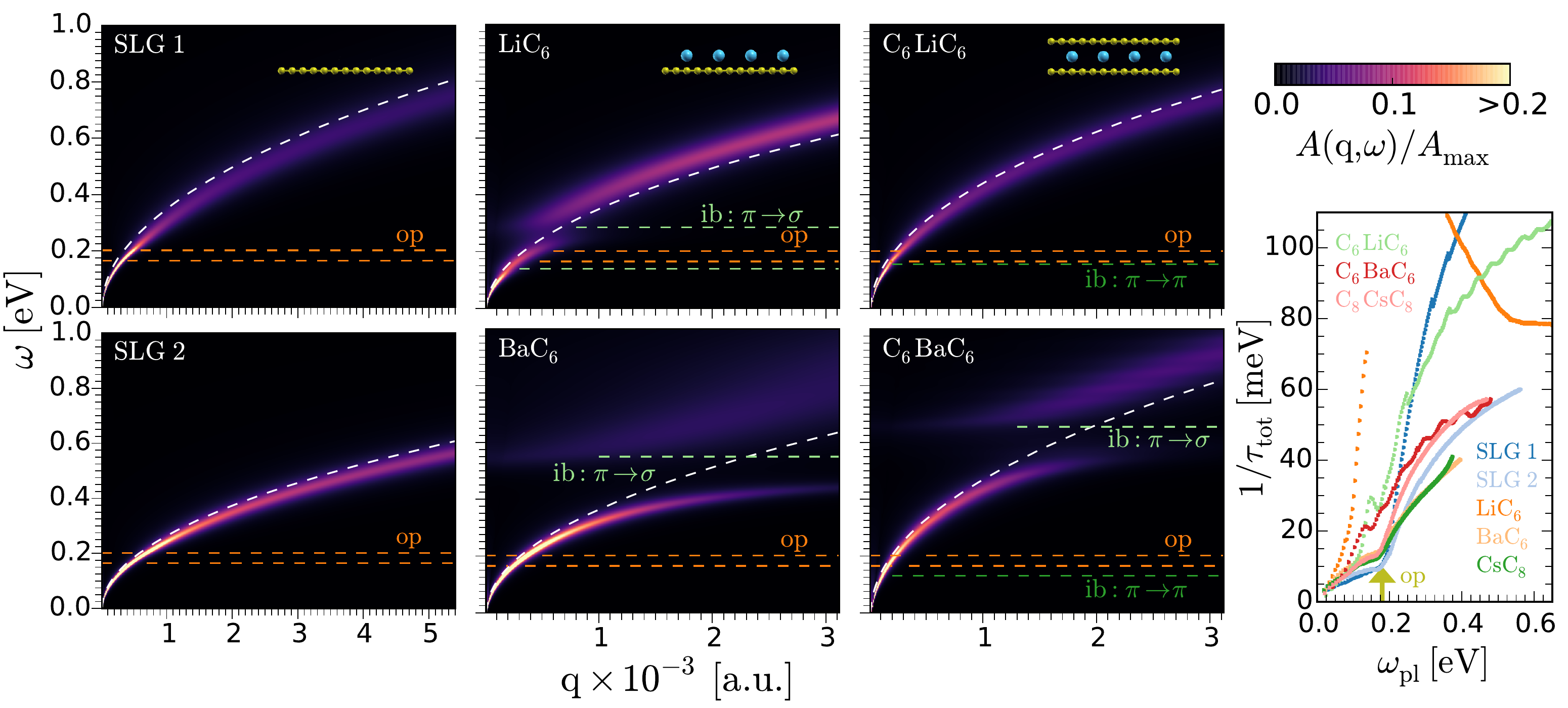}
\caption{\label{fig:fig3}Excitation spectra $A(\mathrm{q},\omega)$ showing plasmon dispersion relations for X-doped single-layer and bilayer graphene. In addition, the corresponding results for SLG--RBA are shown (SLG\,1 has $\varepsilon_F=1.6$\,eV and SLG\,2 has $\varepsilon_F=1.22$\,eV). White dashed lines represent the bare intraband plasmon energies without electron-phonon coupling. Orange dashed lines indicate the boundaries of the C$_{xy}$ optical phonon (op) band. Light and dark green dashed lines show the peak positions of the interband (ib) $\pi\rightarrow\sigma$ and $\pi\rightarrow\pi$ transitions, respectively. On the right the corresponding total plasmon damping rates (i.e., linewidths) $1/\tau_{\mathrm{tot}}$ as a function of the plasmon energy $\omega_{\mathrm{pl}}$ are depicted.}
\end{figure*}

In Fig.\,\ref{fig:fig3} 2D plasmons in graphene are shown with two significant improvements upon the state-of-the-art \textit{ab initio} studies~\cite{bib:vito17,bib:shirodkar17,bib:pisarra14,bib:despoja13,bib:pisarra16,bib:gomez16}: real doping beyond RBA and dynamical phonon-induced damping calculated from first-principles. The SLG--RBA results (SLG\,1 and SLG\,2) clearly show how phonon-induced processes introduce the plasmon linewidth. Even further, the electron-phonon interaction redshifts the plasmon energies (see SI for the corresponding intensities of the plasmon energy renormalization). The impact of the $\pi\rightarrow\sigma$ transitions is equally impressive. For the SLG doped with Li and Ba, and for bilayer graphene doped with Ba, the anticrossing splitting caused by the large hybridization between the 2D plasmon and $\pi\rightarrow\sigma$ excitations is observed (the same holds for Cs, but it is not shown). The coupling is so strong that the usual dispersion of the 2D plasmon is preserved only up to $\omega_{\mathrm{pl}}\approx 0.5\varepsilon_{\pi\rightarrow\sigma}$, where $\varepsilon_{\pi\rightarrow\sigma}$ is the energy of the $\pi\rightarrow\sigma$ excitations. Since $\varepsilon_{\pi\rightarrow\sigma}$ changes as a function of the X dopants and the number of graphene layers, the dispersion relation of the hybrid mode changes as well. These hybrid modes are also highly broadened around and above $\varepsilon_{\pi\rightarrow\sigma}$ due to Landau damping. Note that this hybrid mode bears parallelism with plasmon-plasmon mode investigated in dopant-free bilayer graphene~\cite{bib:low14}, and with plasmon-phonon modes seen in graphene deposited on SiO$_2$~\cite{bib:yan13} or SiC~\cite{bib:koch16} substrates. 

Finally, in Fig.\,\ref{fig:fig3} the results are summarized in the form of the total plasmon decay rates $1/\tau_{\mathrm{tot}}$ as a function of plasmon energy $\omega_{\mathrm{pl}}$. As noted before, in all cases $1/\tau_{\mathrm{tot}}$ is characterized by a sudden increase approximately at the energies of the C$_{xy}$ optical phonon band. However, substantial values of $1/\tau_{\mathrm{tot}}$ are found even below these energies, coming from the low-energy C$_{z}$ and X$_{xy}$ modes or from the interband excitations (e.g., see values of $1/\tau_{\mathrm{tot}}$ around and below 0.2\,eV for $\mathrm{X=Li}$). Authors in Ref.\,\citenum{bib:shirodkar17} use static (dc) phonon-induced decay rate $1/\tau_{\mathrm{ph}}(0)$ coming from the C$_{xy}$ modes to describe plasmon damping in Li-doped bilayer graphene, while the energy dependence of $1/\tau_{\mathrm{tot}}$ only comes from Landau damping. Here I show that using a dc phonon-induced rate and disregarding the C$_{z}$ and X$_{xy}$ modes for estimating plasmon losses is not justified. In fact, they report the value of $23\,\mathrm{meV}$ by considering both electron-phonon and electron-electron scattering channels  [i.e., $1/\tau_{\mathrm{ph}}(0)+1/\tau_{\mathrm{el}}(0)$], while I get an energy-dependent $1/\tau_{\mathrm{ph}}(\omega)$ ranging up to $120\,\mathrm{meV}$. The discrepancy between dc and dynamical values of the decay rate in graphene was also discussed in Refs.\,\cite{bib:fei12,bib:yan13,bib:principi13a}.

With these final results in mind the following optimal systems and the corresponding energy windows with the smallest damping rates can be proposed: Ba- and Cs-doped SLG up to $\approx0.4$\,eV and Ba- and Cs-doped bilayer graphene up to $\approx0.5$\,eV. Conversely, for the same energy window the 2D plasmon in Li-doped graphene undergoes the largest damping due to both electron-phonon coupling and interband excitations. For $\omega\gtrsim0.5$\,eV the ordering is reversed, i.e., Ba and Cs introduce a larger damping than Li (see also SI). I note here that these damping characteristics are specific for
the stoichiometries studied in this work and will vary as a function of concentration of dopants. In fact, higher (lower) concentrations of dopants will introduce higher (lower) concentrations of conducting electrons, which will increase (decrease) the electron-phonon coupling strengths. Additionally, the energy of the $\pi\rightarrow\sigma$ excitations $\varepsilon_{\pi\rightarrow\sigma}$ varies as a function of the concentration of dopants~\cite{bib:vito17}, thus the interband damping channel will be affected as well. Further study of these effects is indeed needed, but goes beyond the scope of the present work.

In conclusion, using first principles calculations on the phonon-induced processes and interband transitions, I have shown that plasmon losses in single-layer and bilayer graphene are largely modified by chemical doping with alkali and alkaline earth metal atoms. In addition to the well-known damping channel that comes from the intrinsic in-plane optical phonons of graphene, here I report two new dopant-associated channels: loss due to low-energy phonon modes (out-of-plane graphene motion and in-plane dopant motion), and loss due to interband excitations between graphene and dopant states. Interestingly, the latter excitation strongly interacts with the 2D plasmon and together they form a new hybrid mode. All these features depend considerably on the dopant species and the number of graphene layers, which opens new possibilities for tuning the 2D plasmons in graphene. These findings may aid future design of graphene-based plasmonic devices in selecting the optimal operation energy range and chemical dopants. Finally, the developed methodology is quite general and can be used for quantitative analysis of phonon-induced plasmon losses in other doped two-dimensional materials.


\begin{acknowledgement}
I gratefully acknowledge F. Caruso, M. Blanco-Rey, C. Draxl, I. Kup\v{c}i\'{c}, and V. Despoja for useful discussions and comments. Financial support by Donostia International Physics Center (DIPC) during various stages of this work is highly acknowledged. Computational resources were provided by the DIPC computing center.
\end{acknowledgement}

\begin{suppinfo}
More informations on crystal and electronic structures, computational details, Eliashberg functions, phonon-induced renormalization and scattering time parameters, and overall damping rates\\
\end{suppinfo}

\bibliography{grplph}

\begin{mcitethebibliography}{49}
\bibitem[Not()]{Note-1}
Note that the experiments obtain larger electron-phonon coupling for Ca than
  Li. The reason could be that Ca atoms are actually intercalated between the
  graphene sheet and Au substrate, which could significantly alter the values
  of $\lambda_{\mathrm{ph}}(0)$~\cite{bib:fedorov14}.\relax
\mciteBstWouldAddEndPunctfalse
\mciteSetBstMidEndSepPunct{\mcitedefaultmidpunct}
{}{\mcitedefaultseppunct}\relax
\EndOfBibitem
\bibitem[Not()]{Note-2}
Another possibility leading to the shift of the $\pi\rightarrow\sigma$ damping
  channel to higher energies would be to change the coverage of X
  atoms~\cite{bib:vito17}\relax
\mciteBstWouldAddEndPuncttrue
\mciteSetBstMidEndSepPunct{\mcitedefaultmidpunct}
{\mcitedefaultendpunct}{\mcitedefaultseppunct}\relax
\EndOfBibitem
\end{mcitethebibliography}

\pagebreak

\begin{table}[t]
\caption{\label{tab:table1}Decomposed $1/\tau_{\mathrm{ph}}(\infty)$ in C and X atoms in-plane ($xy$) and out-of-plane ($z$) contributions for $\mathrm{X=Li}$ and $\mathrm{X=Ba}$. The corresponding results for SLG--RBA are also shown.}
\begin{tabular}{ccccc}
 &  \multicolumn{4}{c}{$1/\tau_{\mathrm{ph}}(\infty)\approx\pi\left\langle \omega\lambda_{\mathrm{ph}}(0) \right\rangle$ [meV]}\\
 & $\mathrm{C}_{xy}$ & $\mathrm{C}_{z}$ & $\mathrm{X}_{xy}$ & $\mathrm{X}_{z}$ \\ \hline
\rule{0pt}{3ex}
$\mathrm{LiC_6}$& 65.62 &  26.14 &  8.74  & 1.48   \\
$\mathrm{BaC_6}$& 33.58 &  4.91  &  2.84  &  1.19   \\
$\mathrm{SLG\,1}$\textsuperscript{\emph{a}}& 178.52 &  5.02 & -   &  -  \\
$\mathrm{SLG\,2}$\textsuperscript{\emph{b}}& 77.26 &  4.23 & -   &  -  \\
\hline
\end{tabular}

  \textsuperscript{\emph{a}} RBA with $\varepsilon_F=1.6\,\mathrm{eV}$;
  \textsuperscript{\emph{b}} RBA with $\varepsilon_F=1.22\,\mathrm{eV}$.
\end{table}
\pagebreak

\end{document}